\documentclass[a4paper,11pt]{article}
\usepackage{pos}

\usepackage{subfigure}
\usepackage{placeins}
\usepackage{braket}

\newcommand{\tvec}[1]{\boldsymbol{#1}}
\newcommand{\mvec}[1]{\vec{\mskip 0.5mu #1}\mskip 1.5mu}
\newcommand{\dd}{\mathrm{d}}
\newcommand{\Op}{\mathcal{O}}
\newcommand{\uup}{u^{\uparrow}}
\newcommand{\dup}{d^{\uparrow}}
\newcommand{\udown}{u^{\downarrow}}
\newcommand{\ddown}{d^{\downarrow}}
\newcommand{\avsum}{\sideset{}{^\prime}\sum}
\newcommand{\ms}{\mskip 1.5mu}
\newcommand{\half}{{\textstyle\frac{1}{2}}}
\newcommand{\tr}{\operatorname{tr}}
\newcommand{\fig}{figure\ }
\newcommand{\Fig}{Figure\ }
\newcommand{\tab}{table\ }
\allowdisplaybreaks

\title{Double parton distributions in the nucleon on the lattice: Flavor interference effects}
\ShortTitle{Nucleon DPDs: Flavor interference}

\author*[a,1]{Christian Zimmermann}
\author[b,1]{Daniel Reitinger}

\affiliation[a]{Aix Marseille Univ, Université de Toulon, CNRS, CPT, Marseille, France}
\affiliation[b]{Universit\"at Regensburg, Universit\"atsstra{\ss}e 31, 93040 Regensburg, Germany}

\note{for the RQCD collaboration}

\emailAdd{christian.zimmermann@univ-amu.fr}
\emailAdd{daniel.reitinger@ur.de}

\abstract{Information about double parton distributions (DPDs) can be obtained by calculating four-point functions on the lattice. We continue our study on the first DPD Mellin moment of the unpolarized
proton by considering interference effects w.r.t. the quark flavor. In our simulation we employ an $n_f = 2 + 1$ ensemble with inverse coupling $\beta = 3.4$, and pseudoscalar masses of $m_\pi = 355~\mathrm{MeV}$ and $m_K = 441~\mathrm{MeV}$. The results are converted to the $\overline{\mathrm{MS}}$-scheme at the scale $\mu = 2~\mathrm{GeV}$. We analyze the dependence of the considered Mellin moments on the quark polarization and compare our results with quark model predictions.}

\FullConference{%
  The 39th International Symposium on Lattice Field Theory (Lattice2022),\\
  8-13 August, 2022 \\
  Bonn, Germany 
}

\begin{document}
\maketitle

\section{Introduction}

In LHC experiments double parton scattering (DPS) contributions are non-negligible. Especially after the high-luminosity upgrade the sensitivity to them will be enhanced. In the last decade, significant progress has been made in studying the theory of DPS and of double parton distributions (DPDs), see e.g.\ \cite{Gaunt:2011xd,Ryskin:2011kk,Blok:2011bu,Diehl:2011yj,Manohar:2012jr}.

The DPS contribution is proportional to the integral over the transverse distance $\tvec{y}$ of the two scattering quarks of a product of DPDs $F_{a_1 a_2}(x_1,x_2,\tvec{y})$:
\vspace*{-0.3cm}

\begin{align}
\int \dd^2\tvec{y} \;
    F_{a_1 a_2}(x_1, x_2, \tvec{y}) \, F_{b_1 b_2}(x_1', x_2', \tvec{y}) \,,
\end{align}
where $x_i$ denote the two longitudinal momentum fractions and $a_i$ labels the quark polarization. Since DPDs are poorly known from experiments, as well as in theory, they are often approximated by:
\vspace*{-0.3cm}

\begin{align}
   \label{eq:dpd-pocket}
F_{a_1 a_2}(x_1, x_2, \tvec{y})
& \overset{?}{=} f_{a_1}(x_1)\, f_{a_2}(x_2) \, G(\tvec{y}) \,.
\end{align}
$f_{a_i}$ are ordinary parton distribution functions (PDFs). This leads to the famous pocket formula \cite{Bartalini:2011jp}:
\vspace*{-0.3cm}

\begin{align}
\sigma_{\mathrm{DPS},ij} = \frac{1}{C} \frac{\sigma_{\mathrm{SPS},i}\  \sigma_{\mathrm{SPS},j}}{\sigma_{\mathrm{eff}}} \,,
\label{eq:dpd-pocket-formula}
\end{align}
where $C$ is a combinatorical factor and the effective cross section $\sigma_{\mathrm{eff}}$ is introduced for dimensional consistency. From the assumption \eqref{eq:dpd-pocket}, one can derive that $\sigma_{\mathrm{eff}}$ it is a fixed constant that does not depend on the DPS process.

Access from first principles to DPDs can be provided by lattice QCD simulations. In the past we published several studies for the case of the pion \cite{Bali:2020mij}, as well as for the proton \cite{Bali:2021gel,Zimmermann:2021deq}. One of our main results was the flavor dependence of the transverse distribution $G(\tvec{y})$ appearing in \eqref{eq:dpd-pocket} in the case of the proton, so that \eqref{eq:dpd-pocket-formula} cannot be fulfilled.

In the current work we want to continue our research regarding DPDs on the lattice by extending our calculation on the contributions from flavor interference. These are usually considered to be suppressed.

\section{Double parton distributions and two-current matrix elements}

The definition of (collinear) DPDs for unpolarized protons is given by the following integral of proton matrix elements of two light-cone currents \cite{Diehl:2011yj}:
\vspace*{-0.3cm}

\begin{align}
\label{eq:dpd-def}
F_{a_1 a_2}(x_1,x_2,\tvec{y})
= 2p^+ \int \dd y^- \int \frac{\dd z^-_1}{2\pi}\, \frac{\dd z^-_2}{2\pi}\,
&      e^{i\ms ( x_1^{} z_1^- + x_2^{} z_2^-)\ms p^+}
\nonumber \\
& \times
 \avsum_\lambda \bra{p,\lambda} \mathcal{O}_{a_1}(y,z_1)\, \mathcal{O}_{a_2}(0,z_2) \ket{p,\lambda}
\,,
\end{align}
including the helicity average $\sum_\lambda^\prime = \frac{1}{2}\sum_\lambda$. This equation involves the light-cone operators:
\vspace*{-0.3cm}

\begin{align}
\label{eq:quark-ops}
\mathcal{O}_{a}(y,z)
&= \left.\bar{q}\left( y - \half z \right)\, \Gamma_{a} \, q^\prime\left( y + \half z \right)
   \right|_{z^+ = y^+_{} = 0,\, \tvec{z} = \tvec{0}}\,,
\end{align}
where the Dirac matrix $\Gamma_a$ selects the quark polarization
\vspace*{-0.3cm}

\begin{align}
  \label{eq:quark-proj}
\Gamma_{(qq^\prime)} & = \half \gamma^+ \,, &
\Gamma_{\Delta (qq^\prime)} &= \half \gamma^+\gamma_5 \,, &
\Gamma_{\delta (qq^\prime)}^j = \half i \sigma^{j+}_{} \gamma_5 \quad (j=1,2) \,,
\end{align}
corresponding to unpolarized, longitudinally polarized and transversely polarized quarks. Applying symmetry arguments, we can give a decomposition of the DPDs $F(x_i,\tvec{y})$ in terms of transversely rotationally invariant functions $f(x_i,y^2)$. For flavor diagonal channels these are given by:
\vspace*{-0.3cm}

\begin{align} \label{eq:invar-dpds}
F_{q_1 q_2}(x_1,x_2, \tvec{y}) &= f_{q_1 q_2}(x_1,x_2, y^2) \,,
\nonumber \\
F_{\Delta q_1 \Delta q_2}(x_1,x_2, \tvec{y})
&= f_{\Delta q_1 \Delta q_2}(x_1,x_2, y^2) \,,
\nonumber \\
F_{\delta q_1 q_2}^{j_1}(x_1,x_2, \tvec{y})
&= \epsilon^{j_1 k} \tvec{y}^k\, m f_{\delta q_1 q_2}(x_1,x_2, y^2) \,,
\nonumber \\
F_{q_1 \delta q_2}^{j_2}(x_1,x_2, \tvec{y})
&= \epsilon^{j_2 k} \tvec{y}^k\, m f_{q_1 \delta q_2}(x_1,x_2, y^2) \,,
\nonumber \\
F_{\delta q_1 \delta q_2}^{j_1 j_2}(x_1,x_2, \tvec{y})
&= \delta^{j_1 j_2} f^{}_{\delta q_1 \delta q_2}(x_1,x_2, y^2)
\nonumber \\
&\quad  + \bigl( 2 \tvec{y}^{j_1} \tvec{y}^{j_2}
         - \delta^{j_1 j_2} \tvec{y}^2 \bigr)\ms
   m^2 f^{\ms t}_{\delta q_1 \delta q_2}(x_1,x_2, y^2) \,.
\end{align}
This is the same for flavor interference. In that case, we indicate the quark flavors belonging to one operator by parentheses. For instance, we denote the unpolarized DPD for $ud$-interference as $F_{(du)(ud)}$ and the corresponding longitudinally polarized DPD as $F_{\Delta(du)\Delta(ud)}$.

DPDs cannot be calculated directly on the lattice, since the operators \eqref{eq:quark-ops} involve light-like quark field separations. Instead, we consider Mellin moments of DPDs, where the light-cone operators are reduced to local operators, which are well understood on the lattice. We denote the corresponding rotationally invariant Mellin moments as:
\vspace*{-0.3cm}

\begin{align}
  \label{eq:skewed-inv-mellin-mom-def}
I_{a_1 a_2}(y^2)
&= \int_{-1}^{1} \dd x_1^{} \int_{-1}^{1} \dd x_2^{} \;
   f_{a_1 a_2}(x_1,x_2,y^2) 
\,.
\end{align}
The matrix elements to be calculated have the following form:
\vspace*{-0.3cm}

\begin{align}
  \label{eq:mat-els}
\avsum_\lambda \bra{p,\lambda} J^{\mu_1 \cdots}_{q_1 q_1^\prime, i_1}(y)\,
              J^{\mu_2 \cdots}_{q_2 q_2^\prime, i_2}(0) \ket{p,\lambda} \,,
\end{align}
where we consider three kinds of local currents:
\vspace*{-0.3cm}

\begin{align}
  \label{eq:local-ops}
J_{qq^\prime, V}^\mu(y) &= \bar{q}(y) \ms \gamma^\mu\ms q^\prime(y) \,,
&
J_{qq^\prime, A}^\mu(y) &= \bar{q}(y) \ms \gamma^\mu \gamma_5\, q^\prime(y) \,,
&
J_{qq^\prime, T}^{\mu\nu}(y) &= \bar{q}(y) \ms \sigma^{\mu\nu} \ms q^\prime(y) \,.
\end{align}
In analogy to \eqref{eq:invar-dpds} we can decompose the matrix elements \eqref{eq:mat-els} in terms of Lorentz invariant functions $A(py,y^2)$, $B(py,y^2)$, $\dots$ and a suitable set of basis tensors, so that there is a one-to-one correspondence to the rotationally invariant DPDs. More details on these decompositions can be found in \cite{Bali:2021gel}. For the leading-twist contributions in \eqref{eq:invar-dpds} we can identify:
\vspace*{-0.3cm}

\begin{align}
\label{eq:skewed-mellin-inv-fct}
I_{a_1 a_2}(y^2)
&= \int_{-\infty}^{\infty} \dd(py)\, A_{a_1 a_2}(py,y^2) \,,
\nonumber \\
I^t_{\delta q \delta q^\prime}(y^2)
&= \int_{-\infty}^{\infty} \dd(py)\, B_{\delta q \delta q^\prime}(py,y^2) \,.
\end{align}
DPDs can be approximated in terms of GPDs by inserting a complete set of eigenstates between the two light-cone operators and assuming that the hadron state with lowest energy dominates. For flavor diagonal contributions this leads to a convolution of two impact parameter distributions:
\vspace*{-0.3cm}

\begin{align}
   \label{eq:dpd-fact}
F_{a_1 a_2}(x_1, x_2, \tvec{y})
& \overset{?}{=} \int \dd^2\tvec{b}\; f_{a_1}(x_1, \tvec{b} + \tvec{y})\,
                      f_{a_2}(x_2, \tvec{b}) \,.
\end{align}
For the flavor interference DPD, we get an analogous expression, involving transition distributions $f_{ud}(x,\tvec{y})$, $f_{du}(x,\tvec{y})$, since the flavor changing operators turn the proton into a neutron. On the level of Lorentz invariant functions $A(py,y^2)$ the convolution turns into an integral of Pauli and Dirac form factors $F_{1,2}(t)$, which we shall consider later in this work.

\section{Lattice simulations}

In order to obtain the two-current matrix elements \eqref{eq:mat-els} on the lattice, we have to evaluate four-point functions, which we define as:
\vspace*{-0.3cm}

\begin{align}
C^{ij,\mvec{p}}_{\mathrm{4pt}}(\mvec{y},t,\tau) 
&:= 
	a^6 \sum_{\mvec{z}^\prime,\mvec{z}} 
	e^{-i\mvec{p}(\mvec{z}^\prime-\mvec{z})}\  
	\left\langle \tr \left\{
		P_+ \mathcal{P}(\mvec{z}^\prime,t)\ J_i(\mvec{y},\tau)\ 
		J_j(\mvec{0},\tau)\ \overline{\mathcal{P}}(\mvec{z},0) 
	\right\} \right\rangle\,, 
\label{eq:4ptdef}
\end{align}
where $\overline{\mathcal{P}}$ and $\mathcal{P}$ are the proton creation and annihilation operators and the sum over $\mvec{z}$, $\mvec{z}^\prime$ together with the subsequent phase projects on a definite proton momentum. The two quark currents are placed on the same Euclidean time slice. $P_+$ projects onto the required state with positive parity.

The local two-current matrix element itself is obtained by calculating the ratio of four-point functions and the proton two-point function:
\vspace*{-0.3cm}

\begin{align}
2V \sqrt{m^2 + \mvec{p}^2} 
	\left. 
		\frac{C^{ij,\mvec{p}}_{\mathrm{4pt}}(\mvec{y},t,\tau)}
		{C^{\mvec{p}}_\mathrm{2pt}(t)} 
	\right|_{0 \ll \tau \ll t} &= 
	\left. 
	\frac{
		\sum_{\lambda\lambda^\prime} 
		\bar{u}^{\lambda^\prime}(p) P_+ u^{\lambda}(p)\  
		\bra{p,\lambda} J_i(y)\ J_j(0) \ket{p,\lambda^\prime}
	}{
		\sum_{\lambda} 
		\bar{u}^{\lambda}(p) P_+ u^{\lambda}(p)
	}\right|_{y^0 = 0} \nonumber\\
&= \avsum_\lambda \bra{p,\lambda} J_{i}(y)\,
              J_{j}(0) \ket{p,\lambda} \,,
\label{eq:4pt-spin-sum}
\end{align}
\begin{figure}
\begin{center}
\includegraphics[scale=0.82]{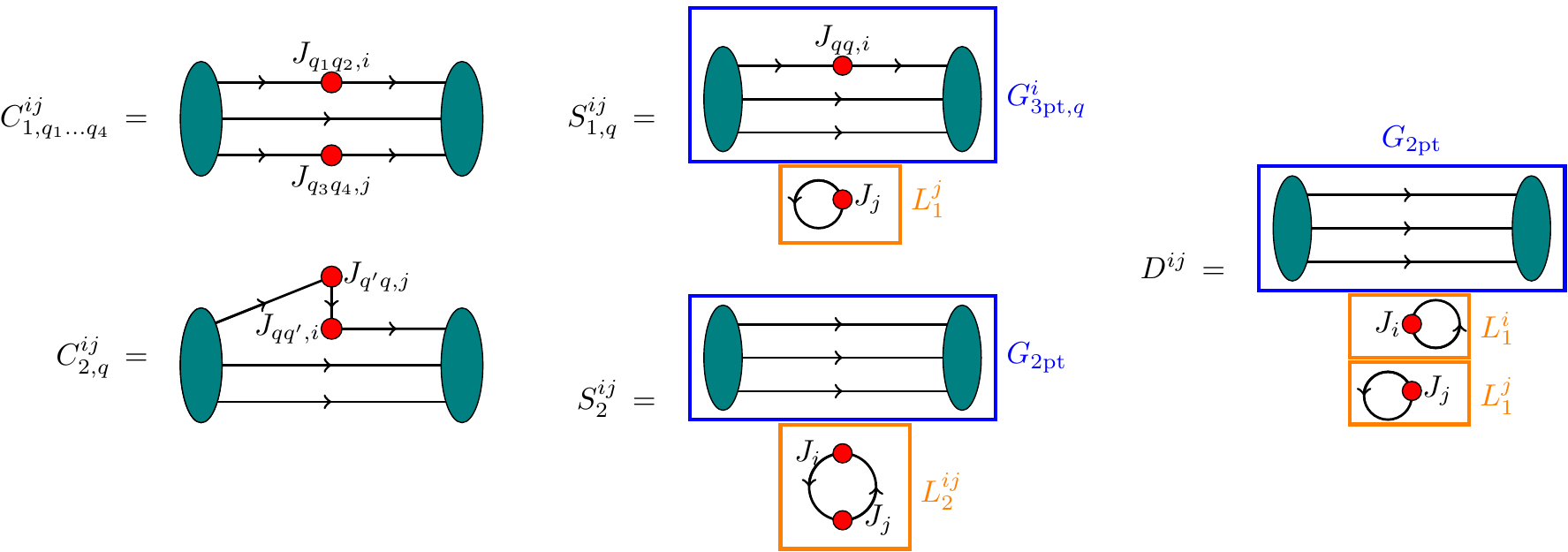}
\end{center}
\caption{Sketch of all types of Wick contractions involved in the calculation of a nucleon four-point function. For the disconnected diagrams we also indicate the disconnected parts, i.e.\ $G_{3\mathrm{pt}}$, $G_{2\mathrm{pt}}$ and the loops $L_1$ and $L_2$. The dark green blobs are the nucleon source and sink, the red points are the insertion operators. \label{fig:graphs}}
\end{figure}
After applying Wick's theorem, the expectation value in \eqref{eq:4ptdef} decomposes in five kinds of Wick contractions. Following the nomenclature of \cite{Bali:2021gel} these are called $C_1$, $C_2$, $S_1$, $S_2$ and $D$. The specific contribution depends on the baryon and the quark flavors of the local operators. In the case of light quarks with equal masses, we have to consider three contributions of $C_1$ topology, namely $C_{1,uudd}$, $C_{1,uuuu}$ and $C_{1,duud}$, where the latter is only relevant for flavor interference contributions. For $C_2$ and $S_1$ there is one contribution for each quark flavor, i.e.\ $C_{2,u}$, $C_{2,d}$, $S_{1,u}$ and $S_{1,d}$, where the flavor is fixed by the quark lines connecting the proton source/sink and one of the insertions. $S_2$ and $D$ are independent of quark flavors for degenerate quark masses. The topology of each contraction kind is sketched in \fig\ref{fig:graphs}. 

Once we fix the quark flavors of the operators we can give an explicit sum of contractions for the considered matrix element. For helicity averaged proton matrix elements we have:

\begin{align}
\left. \bra{p} J_{uu, i}(y)\,
               J_{dd, j}(0) \ket{p}\right|_{y^0 = 0}
&= 
	C^{ij,\mvec{p}}_{1,uudd}(\mvec{y}) + 
	S^{ij,\mvec{p}}_{1,u}(\mvec{y}) + 
	S^{ji,\mvec{p}}_{1,d}(-\mvec{y}) + 
	D^{ij,\mvec{p}}(\mvec{y})\,,
\nonumber \\
\left. \bra{p} J_{uu, i}(y)\,
               J_{uu, j}(0) \ket{p}\right|_{y^0 = 0}
&= 
	C^{ij,\mvec{p}}_{1,uuuu}(\mvec{y}) + 
	C^{ij,\mvec{p}}_{2,u}(\mvec{y}) + 
	C^{ji,\mvec{p}}_{2,u}(-\mvec{y}) + 
	S^{ij,\mvec{p}}_{1,u}(\mvec{y}) + 
	S^{ji,\mvec{p}}_{1,u}(-\mvec{y})
\nonumber \\
&\quad + 
	S_{2}^{ij,\mvec{p}}(\mvec{y}) + 
	D^{ij,\mvec{p}}(\mvec{y})\,,
\nonumber \\
\left. \bra{p} J_{dd, i}(y)\,
               J_{dd, j}(0) \ket{p}\right|_{y^0 = 0}
&= 
	C^{ij,\mvec{p}}_{2,d}(\mvec{y}) + 
	C^{ji,\mvec{p}}_{2,d}(-\mvec{y}) + 
	S^{ij,\mvec{p}}_{1,d}(\mvec{y}) + 
	S^{ji,\mvec{p}}_{1,d}(-\mvec{y}) 
\nonumber \\
&\quad +
	S_{2}^{ij,\mvec{p}}(\mvec{y}) + 
	D^{ij,\mvec{p}}(\mvec{y})\,,
\nonumber \\
\left. \bra{p} J_{du, i}(y)\,
               J_{ud, j}(0) \ket{p}\right|_{y^0 = 0}
&= 
	C^{ij,\mvec{p}}_{1,duud}(\mvec{y}) + 
	C^{ij,\mvec{p}}_{2,u}(\mvec{y}) + 
	C^{ji,\mvec{p}}_{2,d}(-\mvec{y}) + 
	S^{ij,\mvec{p}}_{2}(\mvec{y})\,,
\label{eq:phys_me_decomp}
\end{align}

\begin{table}
\begin{center}
\begin{tabular}{ccccccccccc}
\hline
\hline
id & $\beta$ & $a[\mathrm{fm}]$  & $L^3 \times T$ & $\kappa_{l}$ & $\kappa_{s}$ & $m_{\pi,K}[\mathrm{MeV}]$ & $m_\pi L a$ & configs \\
\hline
H102 & $3.4$ & $0.0856$ & $32^3 \times 96$ & $0.136865$ & $0.136549339$ & $355$, $441$ & $4.9$ & $2037$ \\
\hline
\hline
\end{tabular}
\end{center}
\caption{Details on the CLS ensemble which is used for the calculation of the two-current matrix elements. The simulation includes 990 configurations.\label{tab:cls}}
\end{table}
In our simulation we use a gauge ensemble employing $\Op(a)$-improved $n_f = 2+1$ Sheikholeslami-Wohlert fermions, which has been generated by the CLS collaboration \cite{Bruno:2014jqa}. Details on the ensemble are summarized in \tab\ref{tab:cls}. Our simulation includes 990 configurations. In order to increase the overlap with the ground state, we use boosted nucleon sources and sinks \cite{Bali:2016lva} in combination with APE-smeared gauge fields \cite{Falcioni:1984ei}. The calculation is performed for a fixed nucleon point source, the momentum projection is realized at the sink and by the volume sum at the insertion time slice. In the calculation of the diagrams of type $C_1$, $C_2$ and $S_1$ we make use of the sequential source technique. The contractions of type $C_1$ and $C_2$, as well as the loops $L_1$ of the disconnected diagrams require the usage of stochastic propagators. In the case of $C_2$ and $L_1$ these can be improved using the hopping parameter expansions \cite{Bali:2009hu}. More details on the simulation are given in \cite{Bali:2021gel}

\section{Results}

\begin{figure}
\begin{center}
\subfigure[{\parbox[t]{4cm}{Flavor dependence for $A_{q_1 q_2}$}\label{fig:interf-flav-comp-VV}}]{
\includegraphics[scale=0.24]{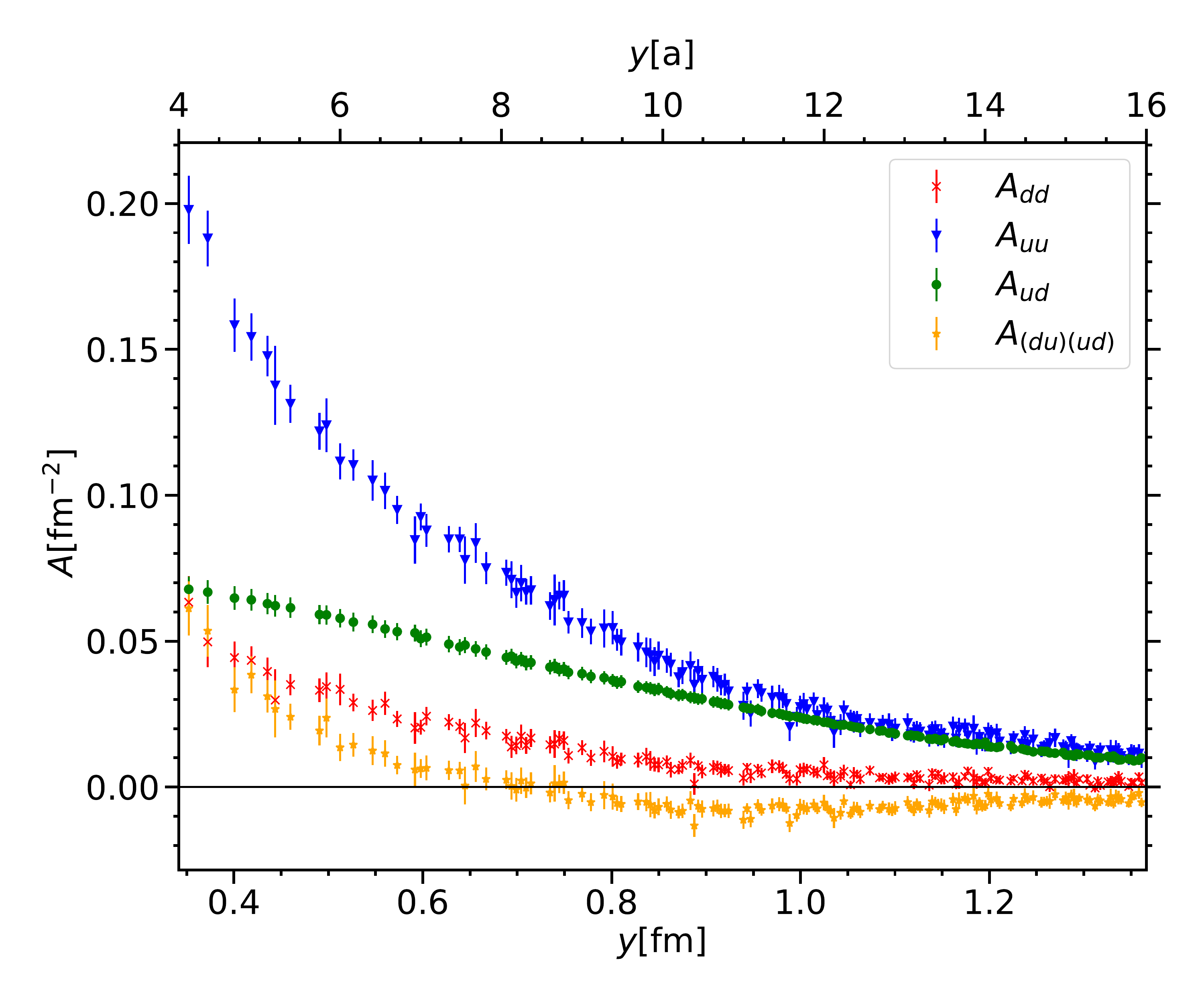}
}
\subfigure[{\parbox[t]{4cm}{Flavor dependence for $A_{q_1 \delta q_2}$}\label{fig:interf-flav-comp-VT}}]{
\includegraphics[scale=0.24]{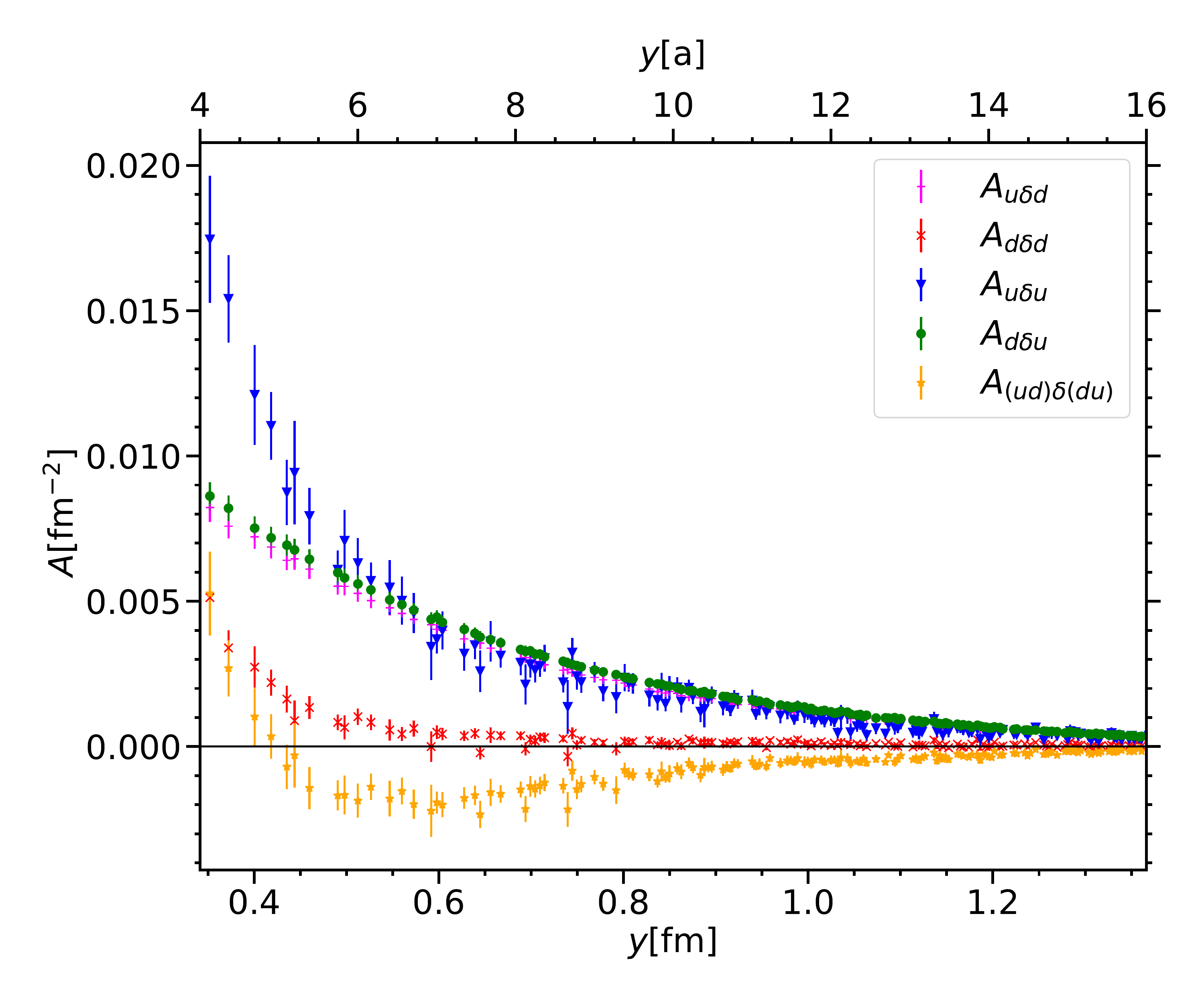}
}
\end{center}
\vspace*{-0.5cm}
\caption{Comparison of the invariant function $A(py,y^2)$ for different quark flavors including the interference contribution (orange). Panel (a) shows the result for two unpolarized quarks, panel (b) those for one unpolarized quark and one transversely polarized quark.\label{fig:interf-flav-comp}}
\end{figure}

In order to estimate the relevance of flavor interference contributions, we compare the corresponding results for given quark polarizations to their flavor diagonal counterparts. This is shown in \fig\ref{fig:interf-flav-comp-VV} for two unpolarized quarks and \fig\ref{fig:interf-flav-comp-VT} for one transversely polarized quark, where the remaining quark is again unpolarized. In both cases we observe values for the interference contributions that have an absolute value comparable to the $dd$ contribution. Moreover, for some values of $y$, the interference signal is negative.

Another interesting aspect to investigate is the comparison of the data obtained from the lattice to predictions by quark models. In this work, we consider a simple $SU(6)$ symmetry w.r.t.\ quark spin and flavor. In this context, the proton wave function can be expressed as:
\vspace*{-0.3cm}

\begin{align}
\ket{p^{\uparrow}} &= 
\frac{1}{3\sqrt{2}} 
\left[ 
\ket{\uup\udown\dup} + \ket{\udown\uup\dup} -2 \ket{\uup\uup\ddown} +
\ket{\uup\dup\udown} + \ket{\udown\dup\uup} -2 \ket{\uup\ddown\uup}
\right. + \nonumber \\
&\qquad + \left.
\ket{\dup\uup\udown} + \ket{\dup\udown\uup} -2 \ket{\ddown\uup\uup} 
\right]\,.
\label{eq:proton-su6}
\end{align}
Although this is a very rudimentary description of the proton neglecting various degrees of freedom like the quark distance, certain quantities like ratios between matrix elements could still be well predicted. Using the $SU(6)$ wave function \eqref{eq:proton-su6}, one can derive the following ratios:
\vspace*{-0.3cm}

\begin{align}
\frac{f_{duud}}{f_{ud}} &= -\frac{1}{2}\,,&\qquad 
\frac{f_{duud}}{f_{uu}} &= -\frac{1}{2}\,,&\qquad 
\frac{f_{ud}}{f_{uu}} &= +1\,, \nonumber\\
\frac{f_{\Delta(du)\Delta(ud)}}{f_{\Delta u \Delta d}} &= -\frac{5}{4}\,,&\qquad 
\frac{f_{\Delta(du)\Delta(ud)}}{f_{\Delta u \Delta u}} &= +\frac{5}{2}\,,&\qquad 
\frac{f_{\Delta u \Delta d}}{f_{\Delta u \Delta u}} &= -2\,, \nonumber\\
\frac{f_{\Delta u \Delta d}}{f_{ud}} &= -\frac{2}{3}\,,&\qquad 
\frac{f_{\Delta u \Delta u}}{f_{uu}} &= +\frac{1}{3}\,,&\qquad 
\frac{f_{\Delta(du)\Delta(ud)}}{f_{duud}} &= -\frac{5}{3}\,.
\end{align}
\begin{figure}
\begin{center}
\subfigure[{\parbox[t]{4cm}{$A_{(du)(ud)}/A_{ud}$ vs $SU(6)$-ratio}\label{fig:su6-ratios-duud-ud}}]{
\includegraphics[scale=0.24, clip, trim=0 0 0 3cm]{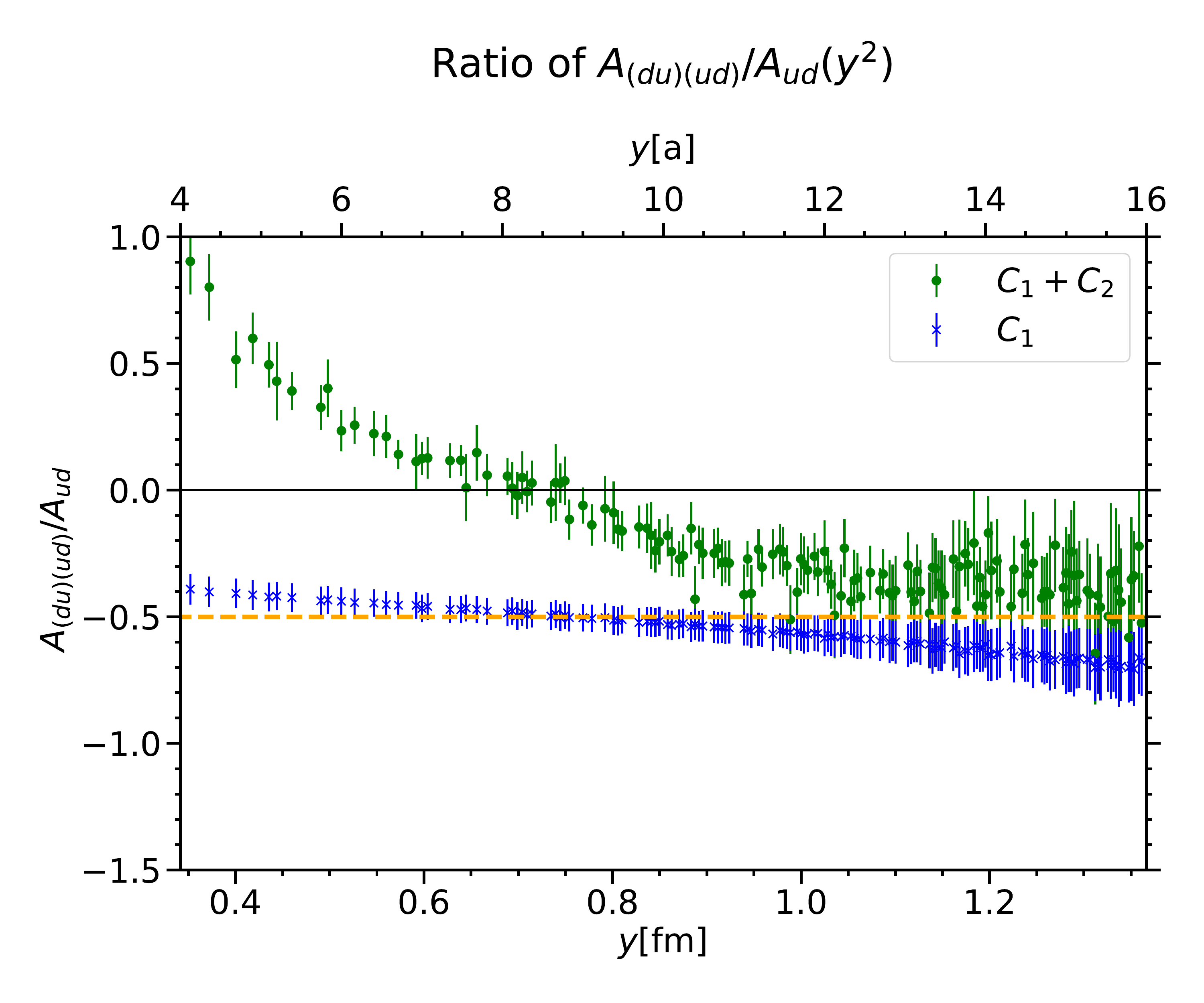}
}
\subfigure[{\parbox[t]{4cm}{$A_{(du)(ud)}/A_{uu}$ vs $SU(6)$-ratio}\label{fig:su6-ratios-duud-uu}}]{
\includegraphics[scale=0.24, clip, trim=0 0 0 3cm]{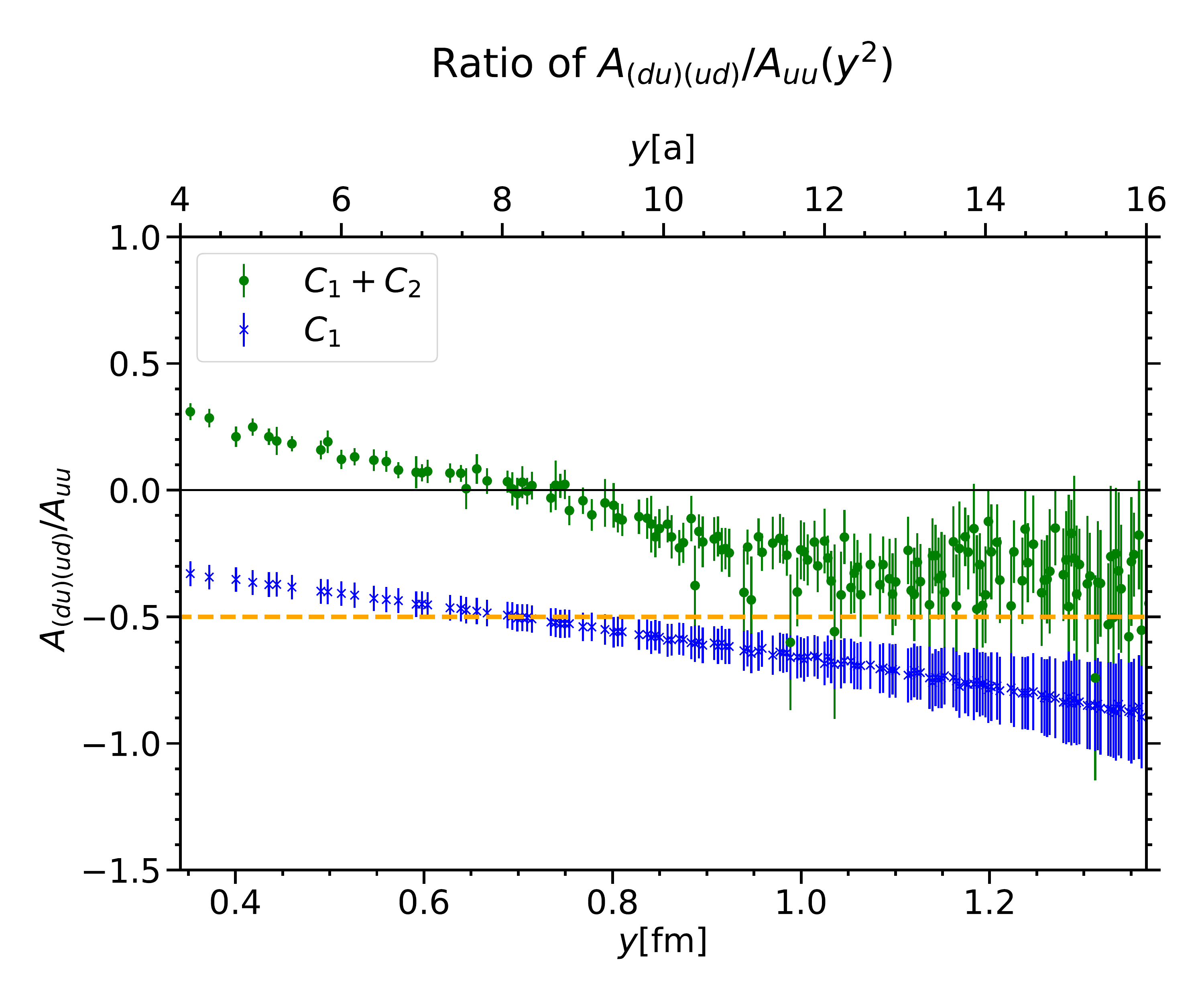}
}
\end{center}
\vspace*{-0.5cm}
\caption{Comparison between ratios of certain flavor channels obtained from the four-point data and the corresponding values predicted by the $SU(6)$ model (dashed, orange line). The blue data points correspond to the result containing $C_1$ data only, while the green data points include all connected diagrams.\label{fig:su6-ratios}}
\end{figure}
In \fig\ref{fig:su6-ratios} we show some selected results for these ratios obtained from our four-point data. Notice that $C_1$ can be considered as more consistent with the picture of a three-quark wave function than $C_2$. Thus, we consider separately the results for the $C_1$ contraction and the result where all connected diagrams are taken into account. For the unpolarized channels, it appears that the $C_1$-data (blue) coincides very well with the $SU(6)$-ratio (orange dashed line). However, for the complete result we observe deviations, especially for smaller $y$. Although not shown, let us note that for the polarized channels the $SU(6)$-ratios do not agree at all with those obtained from the four-point data.

The last aspect we want to consider here is the factorization of DPDs in terms of impact parameter distributions according to \eqref{eq:dpd-fact}. In \cite{Bali:2021gel} we derived the following factorized expression for $A_{qq^\prime}$:
\vspace*{-0.3cm}

\begin{align}
&A_{qq^\prime}(py=0,-\tvec{y}^2) \stackrel{?}{=} 
	\frac{1}{2\pi^2} \int_0^1 \dd \zeta\ 
	\frac{(1-\frac{\zeta}{2})^2}{1-\zeta} 
	\int \dd r\ r J_0 (y r) 
	\left[ 
		K_1(\zeta)\ F_1^q(t)\ F_1^{q^\prime}(t) 
		\vphantom{\frac{\tvec{r}^2}{4m^2}} 
	\right. \nonumber \\
&\qquad\left. 
	- K_2(\zeta) 
		\left( 
			F_1^{q}(t)\ F_2^{q^\prime}(t) + 
			F_1^{q^\prime}(t)\ F_2^{q}(t) 
		\right)
	+ \left( 
		K_3(\zeta) + K_4(\zeta) \frac{\tvec{r}^2}{4m^2} 
	\right) F_2^q(t)\ F_2^{q^\prime}(t)  
\right]\,,
\label{eq:factorization}
\end{align}
where $F_{1,2}^{q}$ are Pauli and Dirac form factors and
\vspace*{-0.3cm}

\begin{align}
K_1(\zeta)~&:=~ 1 - K_2(\zeta)\,, 
&\ 
&K_2(\zeta)~:=~\frac{\zeta^2}{(2-\zeta)^2}\,, 
\nonumber \\
K_3(\zeta)~&:=~\frac{(K_2(\zeta))^2}{K_1(\zeta)}\,, 
&\ 
&K_4(\zeta)~:=~\frac{1}{1-\zeta}\,,
\label{eq:factorization-Ks}
\end{align}
As we already mentioned in the context of \eqref{eq:dpd-fact}, in the case of flavor interference, the usual form factors $F_{i}^{q}$ are replaced by proton-neutron transition form factors $F_{i}^{ud}$ and $F_{i}^{du}$, respectively. In this work, these are calculated from the usual flavor diagonal form factors by exploiting isospin symmetry, so that the involved transition matrix elements can be expressed as:

\begin{align}
\bra{p} J^{ud}_{i} \ket{n} = \bra{p} J^{u}_{i} \ket{p} - \bra{p} J^{d}_{i} \ket{p}
\label{eq:FF-isospin}
\end{align}
\begin{figure}
\begin{center}
\subfigure[{\parbox[t]{4.4cm}{factorization of $A_{ud}(py=0,y^2)$}\label{fig:factorization-ud}}]{
\includegraphics[scale=0.24, clip, trim=0 0 0 3cm]{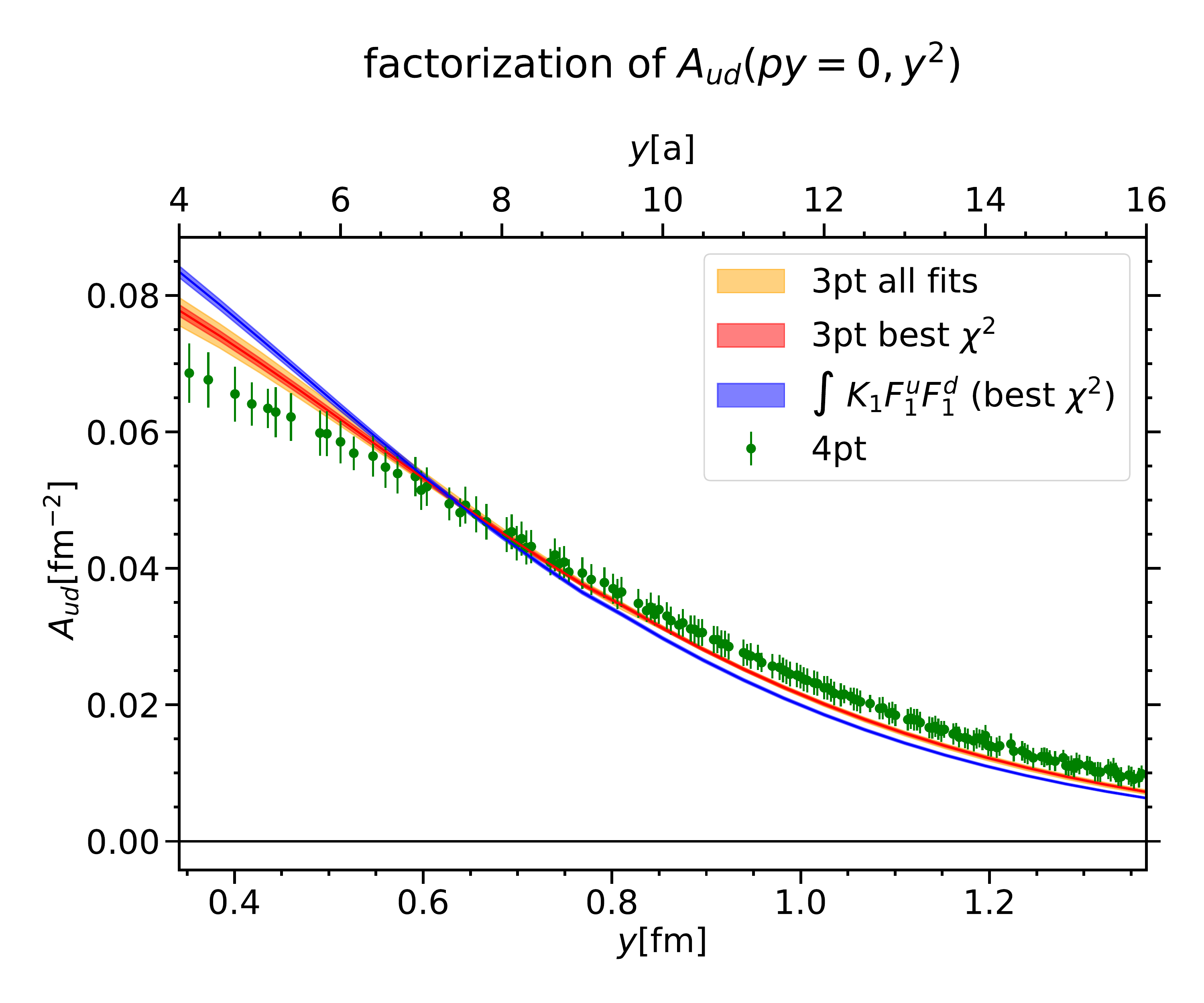}
}
\subfigure[{\parbox[t]{5cm}{factorization of $A_{(ud)(du)}(py=0,y^2)$}\label{fig:factorization-uddu}}]{
\includegraphics[scale=0.24, clip, trim=0 0 0 3cm]{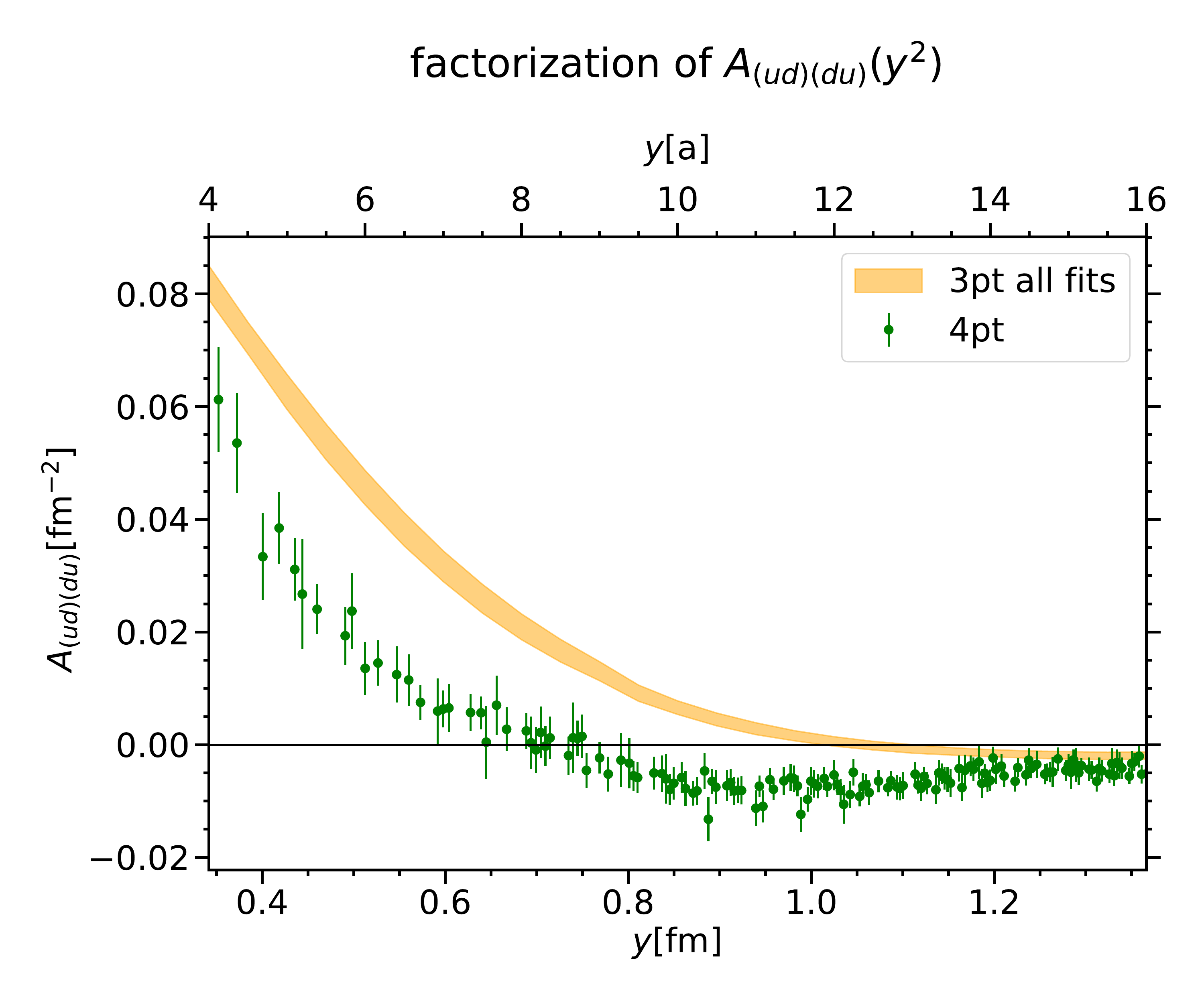}
}
\end{center}
\vspace*{-0.5cm}
\caption{The four-point data (green) is compared to the factorized result obtained from \eqref{eq:factorization}. The left panel shows again the result for $A_{ud}$ given in \cite{Bali:2021gel}. The corresponding test for $A_{(ud)(du)}$ is shown in the right panel. \label{fig:factorization}}
\end{figure}
In order to calculate the integral \eqref{eq:factorization} we use the data of Pauli and Dirac form factors obtained on the lattice in the context of the simulation described in \cite{RQCD:2019jai}.  \Fig\ref{fig:factorization} shows the corresponding results compared to the four-point data (green). In \fig\ref{fig:factorization-ud} we give the result for the flavor diagonal $ud$-channel we already presented in \cite{Bali:2021gel}. An analogous plot for the $uddu$ interference contribution is given by \fig\ref{fig:factorization-uddu}. Again one can observe that the factorization ansatz predicts the correct order of magnitude. However, discrepancies to the four-point data are enhanced.

\section{Conclusions}

We extended our previous analyses of DPDs of the nucleon on the lattice by calculating flavor interference contributions. It was observed that the size of these contributions is comparable to the flavor diagonal $dd$ channel. Moreover, we considered ratios of DPDs for certain flavor combinations which can also be calculated within a $SU(6)$ quark model. On the level of invariant functions, these ratios were compared to the corresponding data obtained on the lattice, where we observed poor agreement, in particular for small $y$. Finally, we checked (again on the level of invariant functions) to what extent the DPDs can be factorized in terms of nucleon form factors. Like for the flavor diagonal channel, we observed that the correct order of magnitude is predicted, however discrepancies are slightly larger. Our next step will be to study the continuum limit.

\section*{Acknowledgments}

We thank the RQCD collaboration, in particular A. Sch\"afer, G. S. Bali  and B. Gl\"a\ss{}le, as well as M. Diehl for fruitful discussions. For providing the proton form factor data, we thank Thomas Wurm. Moreover, we acknowledge the CLS effort for generating the $n_f=2+1$ ensembles, one of which was used for this work. The simulations were performed on the SFB/TRR-55 QPACE 3 cluster. The project leading to this publication received funding from the Excellence Initiative of Aix-Marseille University - A*MIDEX, a French “Investissements d’Avenir” programme, AMX-18-ACE-005.

\end{document}